\documentclass[twocolumn,aps,prl,superscriptaddress,showpacs,floatfix,amsmath]{revtex4-2}
\usepackage{epsfig}
\usepackage{amsmath}
\usepackage{bm}
\usepackage{here}
\usepackage{float}
\usepackage{graphicx}
\usepackage[linkcolor=black,urlcolor=blue,citecolor=blue,colorlinks=true]{hyperref}
\usepackage{xcolor}

\usepackage[normalem]{ulem}

\begin{document}

\title{Unusual magnetic order, field induced melting and role of spin-lattice coupling in 2D Van der Waals materials: a case study of CrSiTe$_3$}

\author{Smita Gohil}
\author{Saswata Halder}
\author{Karthik K Iyer}
\author{Shankar Ghosh}
\author{A. Thamizhavel}
\author{Kalobaran Maiti}
\altaffiliation{email: kbmaiti@tifr.res.in}
\affiliation{Department of Condensed Matter Physics and Materials Science, Tata Institute of Fundamental Research, Homi Bhabha Road, Colaba, Mumbai- 400005}

\date{\today}

%\begin{document}

%\maketitle
\begin{abstract}
Two-dimensional (2D) Van der Waals compounds exhibit interesting electronic and magnetic properties due to complex intra-layer and inter-layer interactions, which are of immense importance in realizing exotic physics as well as advanced technology. Various experimental and theoretical studies led to significantly different ground state properties often contrasting each other. Here, we studied a novel 2D material, CrSiTe$_3$ employing magnetic, specific heat and Raman measurements. Experimental results reveal evidence of incipient antiferromagnetism below 1 kOe concomitant to ferromagnetic order at 33 K. Antiferromagnetic and ferromagnetic interactions coexists at low field in the temperature regime, 15 - 33 K. Low field data reveal an additional magnetic order below 15 K, which melts on application of external magnetic field and remain dark in the heat capacity data. Raman spectra exhibit anomalies at the magnetic transitions; an evidence of strong spin-lattice coupling. Below 15 K, $E_g$ modes exhibit hardening while $A_g$ modes become significantly softer suggesting weakening of the inter-layer coupling at low temperatures which might be a reason for the unusual magnetic ground state and field induced melting of the magnetic order. These results reveal evidence of exceptional ground state properties linked to spin-lattice coupling and also suggest a pathway to study complex magnetism in such technologically important materials.
\end{abstract}

\maketitle

%\section{Introduction:}

Two-dimensional (2D) materials are studied extensively due to their exotic properties important for fundamental science and possibility of on-demand tuning of properties relevant for technological applications \cite{xu_recent_2022, siddique_emerging_2021, sethulakshmi_magnetism_2019, burch_magnetism_2018, Fe-pnictides}. Coexisting semiconducting and magnetic properties aid spintronic applications and can be achieved in these systems via strain engineering, doping, etc. However, such methods often introduces additional complexity due to disorder, defects, etc., which influences the overall band structure, charge carrier mobility, etc. Thus, candidate materials are searched among intrinsic magnetic semiconductors such as transition metal halides, chalcogenides, phosphorus chalcogenides, Mn(Sb)-Bi-Te, Fe$_3$Ge(Ga)Te$_2$, MXY (M = Cr, V; X = O, S, Se, Te; Y= Cl, Br, I) family \cite{wang_magnetic_2022, yang_van_2021, liu2023recent, rahman2021recent, khan2020recent, yang2019van}. These materials retain their magnetic order even if thinned down to a few layers \cite{gong_discovery_2017, liu_critical_2016, lin_ultrathin_2016, zhang2022hard} although 2D geometry does not support long-range order \cite{zhu_topological_2021, mermin1966absence}. In fact, some theoretical studies show enhancement of Curie temperature, $T_c$ with thinning and experimentally a transition from soft to hard ferromagnetism is observed \cite{lin_ultrathin_2016, zhang2022hard}. Evidently magnetism in these materials is complex and lattice appears to play an important role \cite{tian_magneto-elastic_2016, structural-link}.

\begin{figure}
\centering
\includegraphics[width=0.45\textwidth]{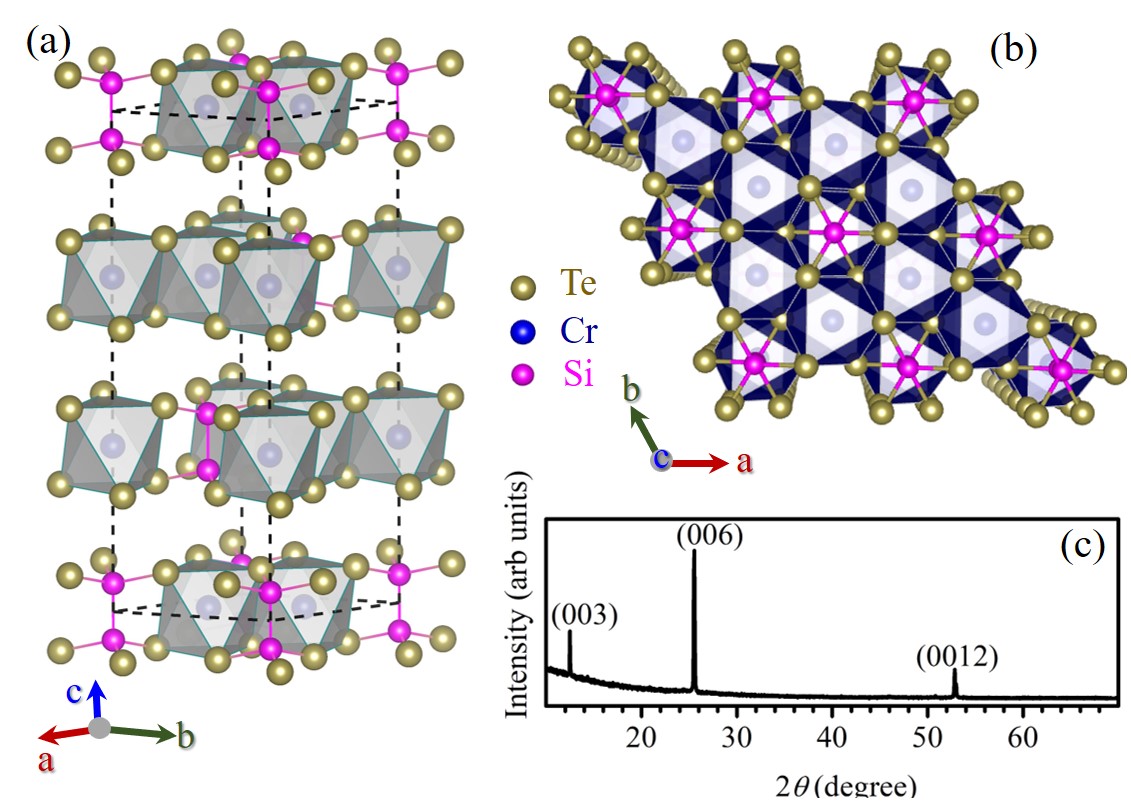}
\caption{(a) Crystal structure of CrSiTe$_3$. (b) Top ($c$-axis) view of the crystal structure exhibiting hexagonal Cr-sublattice with Si-dimers at the center of the hexagon. (c) XRD pattern exhibit single crystallinity and growth along (00$l$)-direction.}
\label{Fig1-Str}
\end{figure}

CrSiTe$_3$ is one such exotic 2D material and forms in a layered rhombohedral structure [$R\overline{3}$ space group (148)] as shown in Fig. \ref{Fig1-Str}(a). CrTe$_6$ octahedra are connected by edge-sharing; Cr-Te-Cr angle close to 90$^o$ favors ferromagnetic coupling according to Goodenough-Kanamori rules \cite{carteaux_2d_1995, zhu_topological_2021, williams2015magnetic}. Si-atoms form dimers in between the Te-planes as shown in the figure and has distorted tetrahedral symmetry. [Te-Si-Cr-Si-Te] quintuple layers are connected by weak Van der Waals force. A top-view of the structure is shown in Fig. \ref{Fig1-Str}(b) exhibiting hexagonal symmetry of the Cr-sites in the $ab$-plane with the Si dimers at the centre.

Various studies showed that CrSiTe$_3$ is an intrinsic semiconductor having ferromagnetic ground state below 33 K \cite{liu_critical_2016, lin_ultrathin_2016}. Electron correlation influences inter-layer exchange interactions leading to antiferromagnetism at higher correlation regime. CrSiTe$_3$ lies at the intersection of ferromagnetic and A-type antiferromagnetic states \cite{zhang_unveiling_2019, li_abnormal_2023, yang_van_2021, sivadas_magnetic_2015, li2019intrinsic}. Experimental verification of this scenario is not reported so far. Employing first principles calculations, Sivadas {\it et al.} found antiferromagnetic order with a zigzag spin texture in CrSiTe$_3$, whereas a similar compound CrGeTe$_3$ exhibits ferromagnetism with $T_c$ = 106 K \cite{sivadas_magnetic_2015}. In the bulk material, the intra-layer magnetic ordering is ferromagnetic. The inter-layer coupling is antiferromagnetic, which becomes ferromagnetic on application of tensile strain along $c$-direction \cite{sivadas_magnetic_2015, sethulakshmi_magnetism_2019}. Large tensile strain ($\geq$ 8\%) in a monolayer enhances $T_c$ to 290 K while the compressive strain leads to antiferromagnetism \cite{siddique_emerging_2021, chen_strain-engineering_2015} suggesting important role of the lattice degrees of freedom in the magnetism of such materials. A recent study reported signature of topological magnon in CrSiTe$_3$ and the emergence of in-gap edge states in the magnon band structure \cite{zhu_topological_2021}. Evidently, CrSiTe$_3$ is exotic posing several outstanding puzzles in the magnetism of such materials. Here, we show that magnetization measured at different magnetic fields on high quality single crystals reveals evidence of incipient antiferromagnetic order along with ferromagnetism. Raman spectra exhibit interesting evolution with temperature manifesting role of spin-lattice coupling in magnetism. Interestingly, experimental data at low temperatures show signature of an additional magnetic order below 15 K and melting of magnetization on application of high magnetic field below 15 K.

%\section{Experimental Details:}

%{\it Sample preparation}:
High quality bulk single crystals of CrSiTe$_3$ were synthesized using self flux growth technique taking Cr (99.999\%), Si (99.999\%) and Te (99.999\%) in the molar ratio of 1:2:6; excess quantity of Si and Te served as self-flux. All these elements were heated together to 1150 $^\circ$C in a sealed quartz tube for 16 hours followed by slow cooling (3~$^\circ$C/h) to 700 $^\circ$C. Extracted single crystals were mirror shiny and the composition was confirmed by
%{\it Characterization}:
energy dispersive analysis of x-rays (EDX) using a Scanning Electron Microscope (FESEM-Zeiss Ultra 55). X-ray diffraction (XRD) pattern collected using PANalytical X'Pert PRO system show (00\textit{l}) family of planes (see Fig. \ref{Fig1-Str}(c)) establishing excellent single crystallinity.

%{\it Measurements}:
Bulk properties measurements were carried out using a Physical Property Measurement System (PPMS) (Quantum Design USA). Magnetic susceptibility ($\chi$) measurements were done using vibrating sample magnetometer (VSM) option of the PPMS.
Raman spectroscopy measurements were carried out using the T64000 Raman spectrometer from Horiba Jobin Vyon. This is a triple grating spectrometer equipped with a liquid nitrogen cooled charged coupled device (CCD) detector. The excitation wavelength used for our studies was 514.5 nm from a tunable gas laser (Stabilite 2018 model, Spectra Physics). Temperature dependent measurements were carried out using continuous flow helium cryostat (MicrostatHe) from Oxford Instruments.

%\end{enumerate}

%\section{Results and Discussion}

\begin{figure}
\centering
\includegraphics[width=0.48\textwidth]{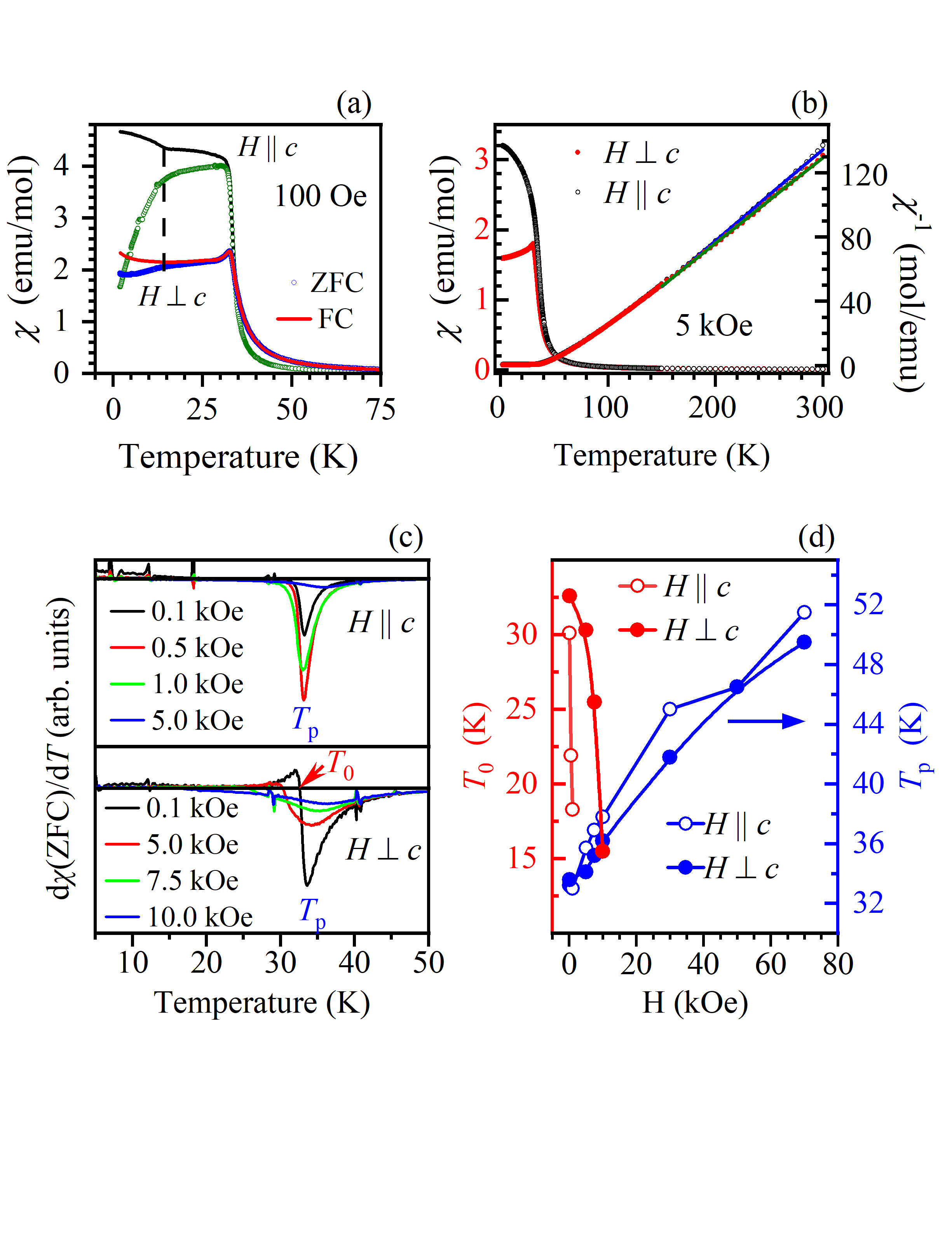}
\vspace{-14ex}
\caption{(a) Magnetic susceptibility, $\chi$ [ZFC (open symbols) and FC (lines)] measured at 100 Oe field parallel and perpendicular to $c$-axis. (b) $\chi$(ZFC) measured at 5 kOe field parallel (open symbols) and perpendicular (closed symbols) to $c$-axis. $\chi^{-1}$ (right-axis) vs. $T$ plots exhibit similar slopes for both the directions. (c) $d\chi/dT$ vs. $T$ at different magnetic fields. Peak, $T_p$ represents Curie temperature. $T_0$ ($d\chi/dT = 0$) represents Ne\'{e}l temperature. (d) $T_p$ and $T_0$ vs. magnetic field parallel (open symbols) and perpendicular (closed symbols) to $c$-axis.}
\label{Fig2-Magn}
\end{figure}

Magnetic susceptibility, $\chi$ measured for field cooled (FC) and zero field cooled (ZFC) conditions at 100 Oe field parallel and perpendicular to the $c$-axis are shown in Fig. \ref{Fig2-Magn}(a). A sharp increase of $\chi$ is observed between 30-35 K suggesting onset of ferromagnetic order consistent with earlier reports \cite{liu_critical_2016, lin_ultrathin_2016}. Experimental data show strong anisotropy in the magnetically ordered state; higher magnetization for $H{\parallel}c$-direction along with significantly sharper rise of $\chi$ compared to the data for the $H{\perp}c$ case. While the FC-ZFC bifurcation near 33 K is large for $H{\parallel}c$, it is almost non-existent in the $H{\perp}c$-case. Above results suggest that the easy axis is along out-of $ab$-plane direction and the magnetic order in $ab$-plane is ferromagnetic.

For $H{\perp}c$, the experimental data exhibit a peak-like feature typical of an antiferromagnetic interaction which is not distinguishable in the case of $H{\parallel}c$. Considering that the electron dynamics due to magnetic field appears in the plane perpendicular to the field direction (Lorentz force), the observation of a distinct peak for $H{\perp}c$ suggests antiferromagnetic inter-layer interactions. Interestingly, there is an additional transition around 15 K exhibiting bifurcation of FC and ZFC data for $H{\perp}c$ and more pronounced bifurcation for $H{\parallel}c$. This was not detected in earlier studies presumably due to the use of high magnetic field for measurements which smeared out relatively weaker couplings. These results indicate emergence of an additional magnetic order at 15 K. Evidently, the magnetism of this system is significantly complex.

In Fig. \ref{Fig2-Magn}(b), we show $\chi(ZFC)$ data measured at 5 kOe field. The differences for the two magnetization directions ($H{\parallel}c$ and $H{\perp}c$) in the temperature range 30 - 50 K as well as signature of multiple transitions observed at 100 Oe field have almost disappeared. The data for $H{\perp}c$ exhibit a peak at around 30 K similar to 100 Oe case and indicates survival of the antiferromagnetic interactions. The $\chi^{-1}$ plots exhibit a linear behaviour in the paramagnetic phase with the similar slopes for both the magnetization directions. This is reflected in the estimated effective magnetic moment of 3.86 $\mu_B$/fu for $H{\parallel}c$ and 3.74 $\mu_B$/fu for $H{\perp}c$ (theoretical value: 3.87 $\mu_B$/fu for Cr$^{3+}$ ion). The paramagnetic Curie temperature, $\theta_p$ is found to be 63 K and 58 K for the $H{\perp}c$ and $H {\parallel}c$, respectively. While positive $\theta_p$ indicates ferromagnetic correlations, $\theta_p$ values are much larger than the observed $T_c$. This suggests possible occurrence of short range correlations as precursors before the compound enters into the magnetically ordered state \cite{precursor}.

To study the magnetic features further, we measured $\chi(ZFC)$ employing different magnetic fields in the temperature regime close to the magnetic transitions - the derivative, $d\chi/dT$ for some cases are shown in Fig. \ref{Fig2-Magn}(c). The narrow $d\chi/dT$ curves for $H{\parallel}c$ suggests sharp transition while it is relatively broader in the $H{\perp}c$ cases. $d\chi/dT$ crosses zero at about 60 K in the high temperature side of the dip, which is close to $\theta_p$. The minima in the $d\chi/dT$ plots, $T_p$ represents the ordering temperature and $T_0$ is the temperature where $d\chi/dT$ becomes zero below $T_p$ \cite{pregelj2010magnetic, sengupta2005magnetic}. Thus, $T_0$ corresponds to the peak in the $\chi$ vs $T$ curves due to antiferromagnetic order, which is most distinct for $H{\perp}c$ data. The estimated $T_0$ and $T_p$ are shown in Fig. \ref{Fig2-Magn}(d). $T_p$ gradually increases with the increase in magnetic field as expected for ferromagnetic interactions and is found to be larger for $H{\parallel}c$ relative to $H{\perp}c$ cases. On the other hand $T_0$ is larger for $H{\perp}c$ and could be identified only in the low field cases. It diminishes with the increase in field strength as expected for antiferromagnetic interactions.

%\subsection{Isothermal Magnetization}

\begin{figure}
\centering
\includegraphics[width=0.48\textwidth]{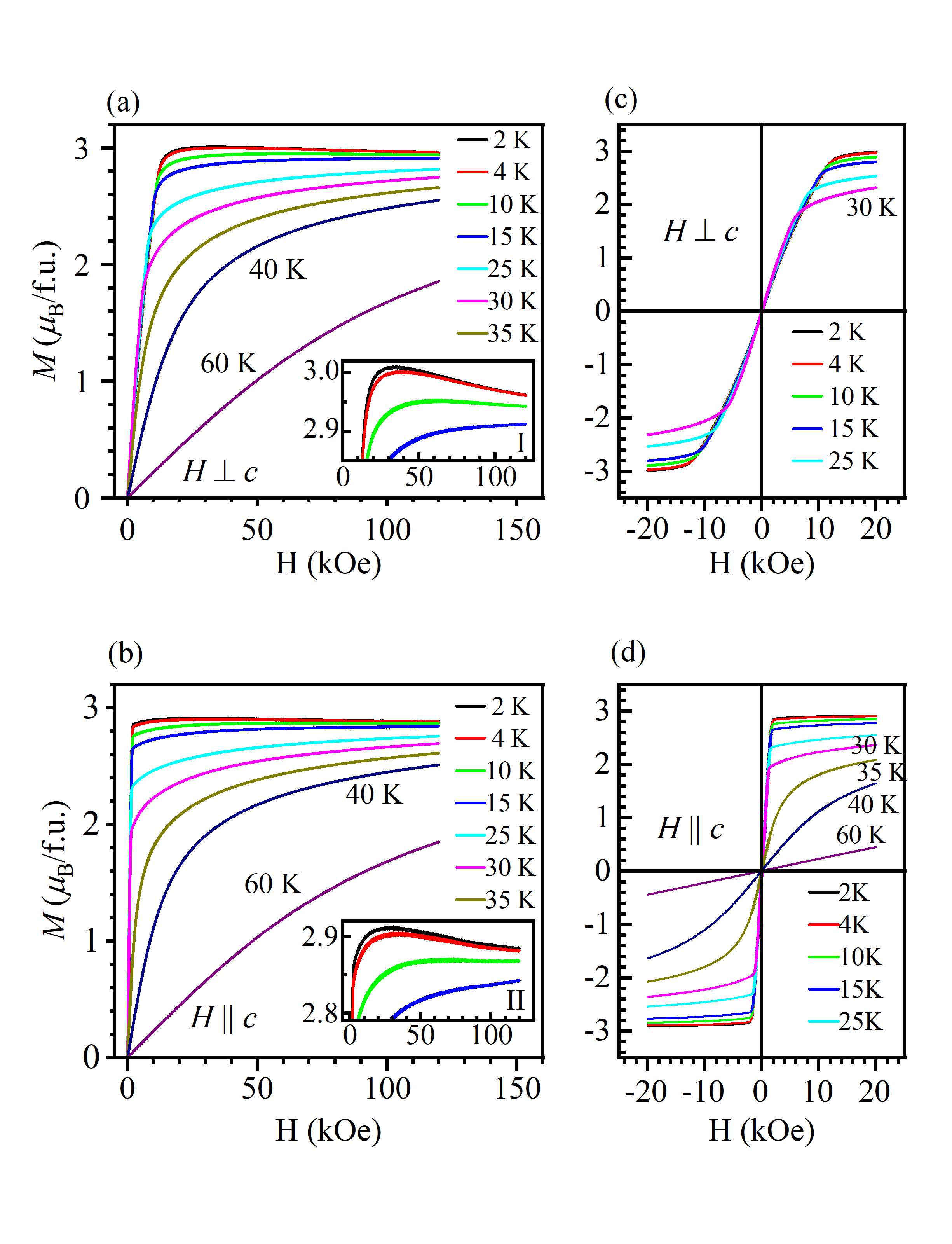}
\vspace{-8ex}
\caption{Isothermal magnetization for (a) $H{\perp}c$ and (b) $H{\parallel}c$. Insets show the same data in an expanded scale manifesting unusual negative slope at high magnetic field below 15 K. Magnetic hysteresis loops at different temperatures for (c) $H{\perp}c$ and (d) $H{\parallel}c$.}
\label{Fig3-MH}
\end{figure}

Isothermal magnetization data as a function of external magnetic fields are shown in Fig. \ref{Fig3-MH}(a) and (b). For $T \leq 15$ K, magnetization data exhibit a sharp increase in response to applied magnetic fields for both the orientations followed by a tendency to saturate. The change of slope for $H{\perp}c$ is seen beyond 10 kOe, while for $H{\parallel}c$, this change is seen at 1 kOe, validating the identification of easy axis along $c$-direction. A close look at the high field magnetization data (see insets I and II) exhibits a decrease in magnetization with increasing magnetic fields beyond about 30 kOe. Magnetization may decrease due to decrease of local magnetic moments. However, such decrease with an increase in field is highly unlikely. The other possibility could be field induced melting of the ordered phase. It has been reported that CrSiTe$_3$ does not exhibit glassy behavior \cite{xie_two_2019}. To verify this below 15 K, we measured the isothermal remanent magnetization, $M_{IRM}$ as a function of time at various temperatures in the magnetically ordered state \cite{mydosh1993spin}. $M_{IRM}$ data do not show relaxation behaviour typical of a glassy system at any of the temperatures studied. For $15 < T \leq 33$ K, the $M H$ data show a monotonic increase of magnetization with the increase in field. The tendency of saturation reduces as the temperature becomes higher. Anomalies observed at $T \leq 15$ K are not present in this temperature regime. For $T > 33$ K, the overall behavior of $MH$ plots remain similar with decreased tendency of saturation as expected due to thermal effect. In Fig. \ref{Fig3-MH}(c) and (d), we show the magnetic hysteresis loops measured at low fields. The $MH$ plots do not show finite loop area similar to soft-ferromagnetic systems.

%\subsection{Heat Capacity}

\begin{figure}
\centering
\includegraphics[width=0.48\textwidth]{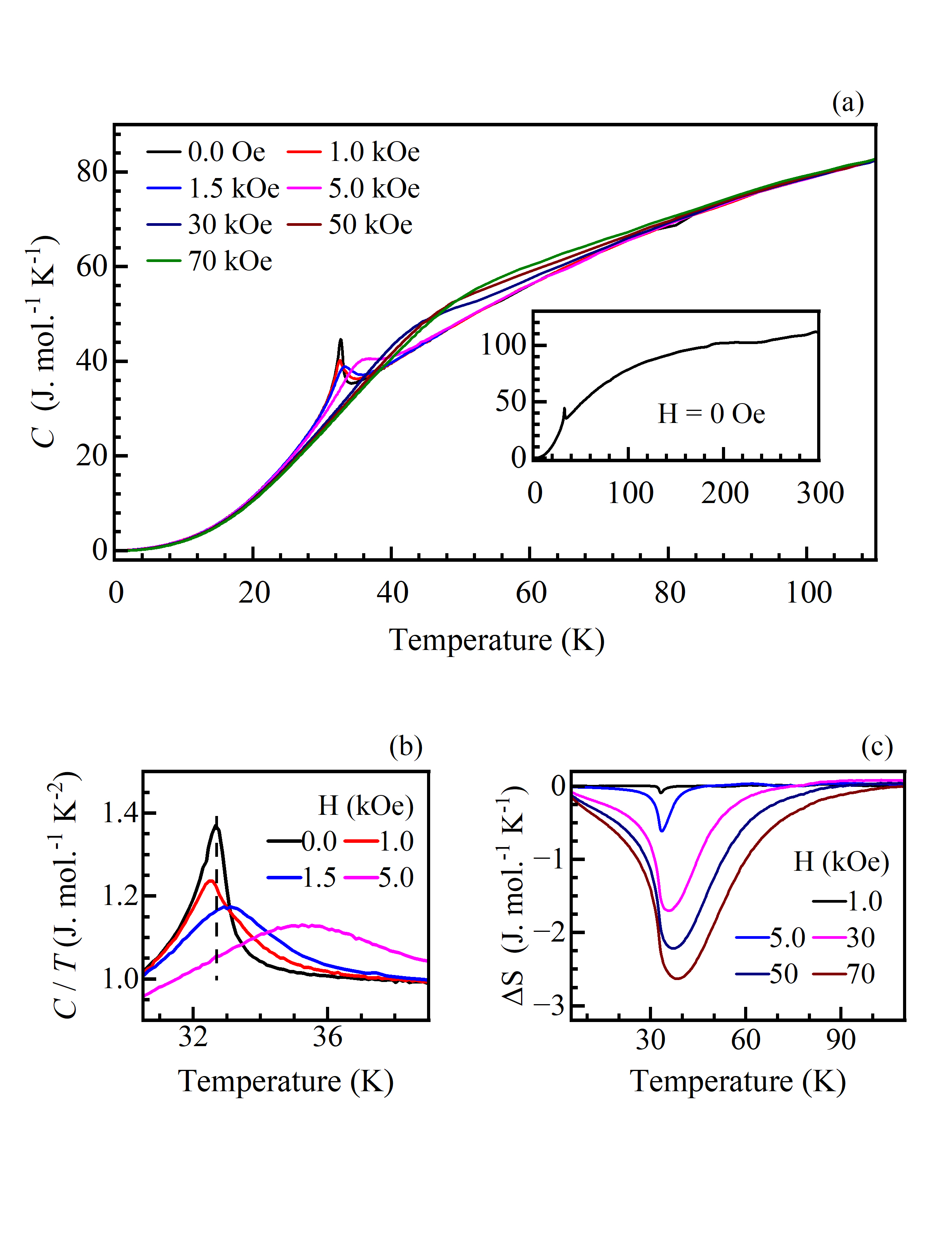}
\vspace{-12ex}
\caption{Heat capacity at different magnetic fields. Inset: zero field data exhibiting close to 3$NR$ value at 300 K. (b) $C/T$ plots near $T_c$. (c) Entropy, $\Delta{S}$ at different magnetic fields.}
\label{Fig4-SpHeat}
\end{figure}

Heat capacity, $C$ measured as a function of temperature is shown in Fig. \ref{Fig4-SpHeat}. Experimental data show large specific heat vanishing at low temperature as expected unlike other insulating systems exhibiting finite value due to disorder \cite{manganite-spHeat}. At zero magnetic field, the overall behavior is similar to typical Debye like temperature dependence \cite{Ayanesh-SnTe} along with a sharp peak ($\lambda$-like anomaly) around 33 K due to magnetic phase transition consistent with the magnetization data. The data exhibits a saturation to about 115 J.mol$^{-1}$.K$^{-1}$ as shown in the inset of Fig. \ref{Fig4-SpHeat}(a), which is close to the classical limit of 3$NR$ for this material. With the application of a magnetic field of 1 kOe, this feature broadens and marginally shifts to lower temperature suggesting the presence of antiferromagnetic interactions as seen in the magnetization data. This scenario is better represented in the $C/T$ versus $T$ plots shown in Fig. \ref{Fig4-SpHeat}(b) \cite{sengupta2005magnetic, kumar2021competing}. Further increase in magnetic field leads to a shift of the peak towards higher temperatures as expected in a ferromagnetic system; the gradual enhancement of the width of the feature is attributed to the thermal disorder. We calculated the isothermal entropy change, $\Delta S$ [= $S(H)-S(0)$] from the $C/T$ vs $T$ data. The estimated values are shown in Fig. \ref{Fig4-SpHeat}(c). For a field of 1 kOe, $\Delta{S}$ shows a weak positive increase due to antiferromagnetism followed by a change in slope with a peak at 33 K in the negative quadrant for the ferromagnetic order. At higher magnetic fields, $\Delta S$ remains in the negative quadrant. We do not observe any other features due to magnetic anomalies at lower temperatures, which suggests that the change in heat capacity due to the magnetic order may be compensated by the changes in the lattice degrees of freedom.

%\subsection{Raman Spectroscopy}

\begin{figure}
\centering
\includegraphics[width=0.48\textwidth]{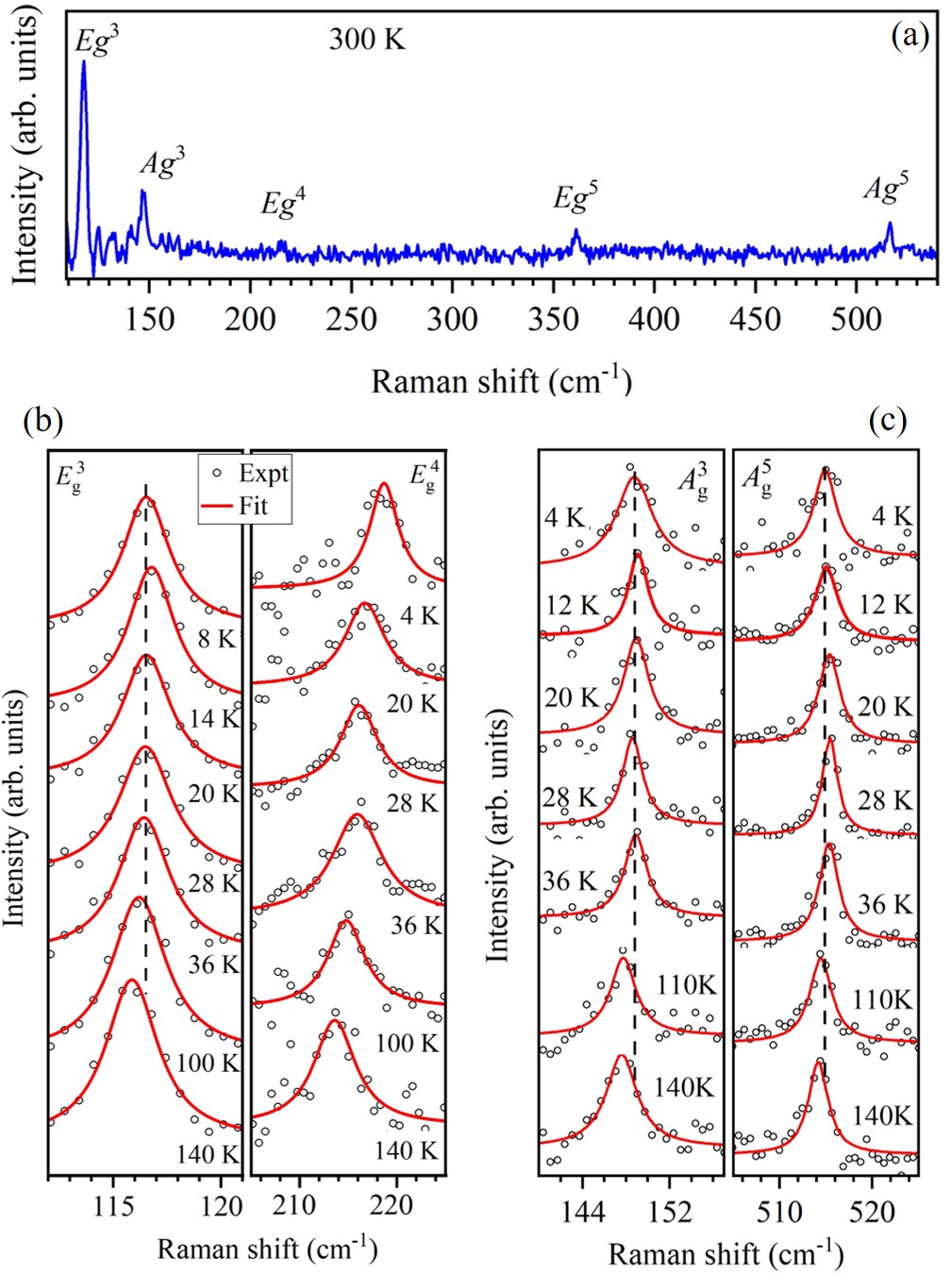}
%\vspace{-18ex}
\caption{(a) Raman spectrum at 300 K. Temperatures evolution of (b) $E_g^3$, $E_g^4$ modes, and (c) $A_g^3$, $A_g^5$ modes. Lines are the Lorentzian fit.}
\label{Fig5-Raman}
\end{figure}

To study the role of the lattice degrees of freedom on magnetic ordering, we performed Raman scattering experiments at varied temperatures down to 4 K on freshly cleaved samples and at low laser power to avoid formation of TeO$_2$ due to burning of the sample\cite{milosavljevic_evidence_2018}. Irreducible representations \cite{milosavljevic_evidence_2018, sun2018effects} of the $\Gamma$-point phonon in $R\overline{3}$ crystal structure are, $\Gamma_{Raman} = 5A_g + 5E_g$; $\Gamma_{IR} = 4A_u + 4E_u$; $\Gamma_{Acoustic} = A_u + E_u$. Since the $A_g$ modes are polarization dependent, we used a circular polarizer in the incident path to capture both $E_g$ and $A_g$ modes. Experimental data at 300 K is shown in Fig. \ref{Fig5-Raman}(a) exhibiting $E_g^3$ (116 cm$^{-1}$), $A_g^3$ (146 cm$^{-1}$), $E_g^4$  (213 cm$^{-1}$), $E_g^5$ (356 cm$^{-1}$) and $A_g^5$ (515 cm$^{-1}$) modes. We studied the evolution of the Raman shift with temperature down to 4 K. In Fig. \ref{Fig5-Raman}(b) and (c), we show the features at few selected temperatures for $E_g$ and $A_g$ modes, respectively exhibiting anomalous shift of the peaks with temperature. The dashed vertical lines in the left panels of Figs. \ref{Fig5-Raman}(b) and (c) show the higher energy of 12 K and 14 K peaks than those at lower and higher temperatures.

\begin{figure}
\centering
\includegraphics[width=0.48\textwidth]{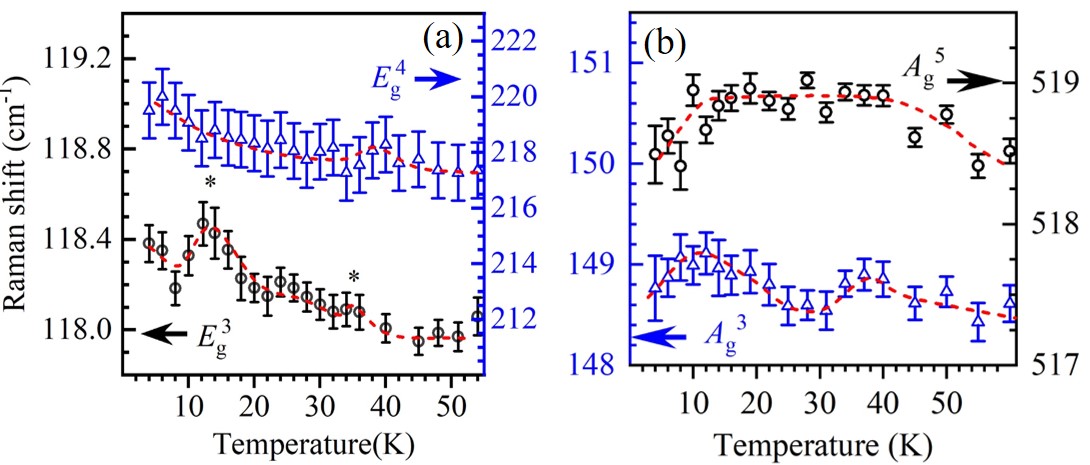}
%\vspace{-18ex}
\caption{Temperature evolution of the (a) $E_g$-modes and (b) $A_g$-modes exhibiting anomalies at magnetic transitions.}
\label{Fig5-RamanTemp}
\end{figure}

Peak positions are estimated by fitting the experimental data with a Lorentzian following least square error method. The fit results (lines) are superimposed over the experimental data in figure. The temperature evolution of the extracted peak positions for $E_g$ and $A_g$ modes are shown in Fig. \ref{Fig5-RamanTemp}(a) and (b), respectively; dashed lines drawn over the experimental data are guides to the eyes. $E_g^4$-mode exhibit weak and gradual shift of the peaks to higher energies indicating hardening of the bonds down to about 15 K with weak anomaly near 33 K. Below 15 K, the hardening becomes faster. $E_g^3$-mode exhibits interesting changes with discernible peaks at about 15 K and 33 K, where the magnetic order sets in. The peak near 15 K is the most intense peak. On the other hand, $A_g$-modes exhibit softening of the bonds below 15 K suggesting weakening of interlayer coupling. $A_g^5$ mode exhibits change in slope instead of peaks; the transition regime is unusually flat. Infrared spectroscopy and lattice dynamic calculations reported signature of a spin lattice coupling via Si-Te stretching mode in CrSiTe$_3$ \cite{casto_strong_2015, sun_review_2021, lu_lattice_2016}. Our results manifest intimate link of spin-lattice coupling and magnetism in this system.

$E_g$-modes involve stretching without significant change in Cr-Te layer separation. Such vibrations of the charge centers will modulate the Cr-Te hybridization and spin-orbit coupling significantly in addition to a change in Cr-Te-Cr bond angle which influences inter-site exchange interactions. On the other hand, $A_g$-vibrations influence inter-layer separation. Softening of these modes below 15 K suggests weakening of the interlayer coupling, which may be a reason for melting of the magnetic order on application of external magnetic field.

%\section{Conclusions}

From the above analysis, a picture of rich phase diagram emerges [see Fig. \ref{Fig2-Magn}(d)]. (i) CrSiTe$_3$ orders ferromagnetically at $T_c \approx$ 33 K and ${\theta_p}\gg{T_c}$ suggests emergence of short-range order as a precursor of the ordered state. Inter-layer coupling is antiferromagnetic for $T <$ 33 K providing an experimental evidence of such interactions predicted earlier. Antiferromagnetism disappears at fields higher than 1 kOe. (ii) For 33 K $\geq T \geq$ 15 K, antiferromagnetic and ferromagnetic interactions co-exists at low magnetic field. (iii) An additional ferromagnetic-type order is discovered at 15 K which could not be captured in heat capacity data; signature of this transition disappears at higher magnetic fields. (iv) Below 15 K, field induced-melting of magnetic order is observed. Evolution of Raman shift with temperature exhibits distinct peaks and/or changes in the slopes revealing structural anomaly at the magnetic transitions. This suggests complex spin-lattice coupling as an underlying mechanism for exotic magnetism as also observed in other systems such as Fe-based superconductors \cite{Fe-pnictides}. Anomalies appear to be most significant below 15 K; in-plane vibrational modes show hardening while the out-of-plane modes becomes softer, which maybe a reason for melting of long-range order. This study opens up a pathway to study complex magnetism in 2D VdW systems involving exceptional spin-lattice coupling, field dependent magnetic interactions, and field-induced melting of long-range order.

%\section{Acknowledgements}
Authors acknowledge the financial support from the Department of Atomic Energy (DAE), Govt. of India (Project Identification no. RTI4003, DAE OM no. 1303/2/2019/R\&D-II/DAE/2079 dated 11.02.2020). K. M. acknowledges financial support from BRNS, DAE, Govt. of India under the DAE-SRC-OI Award (grant no. 21/08/2015-BRNS/10977).

%\end{document}


\begin{thebibliography}{99}

\bibitem{xu_recent_2022}
%title = {Recent advances in two-dimensional van der {Waals} magnets},
H. Xu, S. Xu, X. Xu, J. Zhuang, W. Hao, and Yi Du,
Microstructures \href{https://doi.org/10.20517/microstructures.2022.02}{\textbf{2}, 2022011 (2022)}.


\bibitem{siddique_emerging_2021}
%	title = {Emerging two-dimensional tellurides},
S. Siddique, C. G. Chowde, S. Demiss, R. Tromer, S. Paul, \emph{et al}.,
%K. K. Sadasivuni, E. F. Olu, A. Chandra, V. Kochat, D. S. Galvão, P. Kumbhakar, R. Mishra, P. M. Ajayan, S. C. Tiwary,
Materials Today \href{https://doi.org/10.1016/j.mattod.2021.08.008}{\textbf{51}, 402 (2021)}.


\bibitem{sethulakshmi_magnetism_2019}
%title = {Magnetism in two-dimensional materials beyond graphene},
N. Sethulakshmi, A. Mishra, P. M. Ajayan, Y. Kawazoe, A. K. Roy, \emph{et al}.,
%A. K. Singh, and C. S. Tiwary,
Materials Today \href{https://doi.org/10.1016/j.mattod.2019.03.015}{\textbf{27}, 107 (2019)}.


\bibitem{burch_magnetism_2018}
%title = {Magnetism in two-dimensional van der {Waals} materials},
K. S. Burch, D. Mandrus, and J.-G. Park,
Nature \href{https://doi.org/10.1038/s41586-018-0631-z}{\texttt{563}, 47 (2018)}.


\bibitem{Fe-pnictides}
G. Adhikary, N. Sahadev, D. Biswas, R. Bindu, N. Kumar, \emph{et al}.,
%A. Thamizhavel, S. K. Dhar, and K. Maiti,
%Electronic structure of EuFe$_2$As$_2$
J. Phys.: Condens. Matter \href{https://doi.org/10.1088/0953-8984/25/22/225701}{\textbf{25}, 225701 (2013)};
%
K. Maiti, Pramana - J. Phys. \href{https://doi.org/10.1007/s12043-015-0992-x}{\textbf{84}, 947 (2015)};
% title={Hidden phase in parent Fe-pnictide superconductors}
K. Ali, G. Adhikary, S. Thakur, S. Patil, S. K. Mahatha, \emph{et al}.,
%A. Thamizhavel, G. De Ninno, P. Moras, P. M. Sheverdyaeva, C. Carbone, L. Petaccia, and K. Maiti1,
Phys. Rev. B \href{https://doi.org/10.1103/PhysRevB.97.054505}{\textbf{97}, 054505 (2018)};
%
% title={ Dimensionality, nematicity and superconductivity in Fe-based systems}
K. Ali and K. Maiti,
Eur. Phys. J. B \href{https://doi.org/10.1140/epjb/e2018-90359-2}{\textbf{91}, 199 (2018)}.


\bibitem{wang_magnetic_2022}
%title = {The {Magnetic} {Genome} of {Two}-{Dimensional} van der {Waals} {Materials}},
Q. H. Wang, A. B.-Pinto, M. Blei, A. H. Dismukes, A. Hamo \emph{et al}.,
%S. Jenkins, M. Koperski, Y. Liu, Q.-C. Sun, E. J. Telford,
%Kim, Hyun Ho and Augustin, Mathias and Vool, Uri and Yin, Jia-Xin and Li, Lu Hua and Falin, Alexey and Dean, Cory R. and Casanova, Fèlix and Evans, Richard F. L. and Chshiev, Mairbek and Mishchenko, Artem and Petrovic, Cedomir and He, Rui and Zhao, Liuyan and Tsen, Adam W. and Gerardot, Brian D. and Brotons-Gisbert, Mauro and Guguchia, Zurab and Roy, Xavier and Tongay, Sefaattin and Wang, Ziwei and Hasan, M. Zahid and Wrachtrup, Joerg and Yacoby, Amir and Fert, Albert and Parkin, Stuart and Novoselov, Kostya S. and Dai, Pengcheng and Balicas, Luis and Santos, Elton J. G.,
ACS Nano \href{https://doi.org/10.1021/acsnano.1c09150}{\textbf{16}, 6960 (2022)}.


\bibitem{yang_van_2021}
%title = {van der {Waals} {Magnets}: {Material} {Family}, {Detection} and {Modulation} of {Magnetism}, and {Perspective} in {Spintronics}},
S. Yang, T. Zhang, and C. Jiang,
Adv. Sci. \href{https://doi.org/10.1002/advs.202002488}{\textbf{8}, 2002488 (2021)}.


\bibitem{liu2023recent}
%  title={Recent advances in 2D van der Waals magnets: Detection, modulation, and applications},
P. Liu, Y. Zhang, K. Li, Y. Li, and Y. Pu,
Iscience \href{https://doi.org/10.1016/j.isci.2023.107584}{\textbf{26}, 107584 (2023)}.


\bibitem{rahman2021recent}
%  title={Recent developments in van der Waals antiferromagnetic 2D materials: Synthesis, characterization, and device implementation},
S. Rahman, J. F. Torres, A. R. Khan, and Y. Lu,
ACS nano \href{https://doi.org/10.1021/acsnano.1c06864}{\textbf{15}, 17175 (2021)}.

\bibitem{khan2020recent}
%  title={Recent breakthroughs in two-dimensional van der Waals magnetic materials and emerging applications},
Y. Khan, Sk Md Obaidulla, M. R. Habib, A. Gayen, T. Liang, \emph{et al}.,
%X. Wang, and M. Xu,
Nano Today \href{https://doi.org/10.1016/j.nantod.2020.100902}{\textbf{34}, 100902 (2020)}.

\bibitem{yang2019van}
%  title={Van der Waals engineering of magnetism},
J.-H. Yang, and H. Xiang,
Nat. Mater. \href{https://doi.org/10.1038/s41563-019-0505-2}{\textbf{18}, 1273 (2019)}.

\bibitem{gong_discovery_2017}
%title = {Discovery of intrinsic ferromagnetism in two-dimensional van der {Waals} crystals},
C. Gong, L. Li, Z. Li, H. Ji, A. Stern \emph{et al.},
%Y. Xia, T. Cao, W. Bao, C. Wang, Y. Wang, Z. Q. Qiu, R. J. Cava, S. G. Louie, J. Xia, and X. Zhang,
Nature \href{https://doi.org/10.1038/nature22060}{\textbf{546}, 265 (2017)}.


\bibitem{liu_critical_2016}
%title = {Critical behavior of the quasi-two-dimensional semiconducting ferromagnet {CrSiTe3}},
B. Liu, Y. Zou, L. Zhang, S. Zhou, Z. Wang, \emph{et al}.,
%W. Wang, Z. Qu, and Y. Zhang,
Sci. Rep. \href{https://doi.org/10.1038/srep33873}{\textbf{6}, 33873 (2016)}.


\bibitem{lin_ultrathin_2016}
%title = {Ultrathin nanosheets of {CrSiTe} $_{\textrm{3}}$ : a semiconducting two-dimensional ferromagnetic material},
M.-W. Lin, H. L. Zhuang, J. Yan, T. Z. Ward, A. A. Puretzky, \emph{et al}.,
%C. M. Rouleau, Z. Gai, L. Liang, V. Meunier, B. G. Sumpter, P. Ganesh, P. R. C. Kent, D. B. Geohegan, D. G. Mandrus, and K. Xiao,
J. Mater. Chem. C \href{https://doi.org/10.1039/C5TC03463A}{\textbf{4}, 315 (2016)}.


\bibitem{zhang2022hard}
%  title={Hard ferromagnetic behavior in atomically thin CrSiTe 3 flakes},
C. Zhang, L. Wang, Y. Gu, X. Zhang, X. Xia, \emph{et al}.,
%S. Jiang, L.-L. Huang, Y. Fu, C. Liu, J. Lin \emph{et al.},
Nanoscale \href{https://doi.org/10.1039/D2NR00331G}{\textbf{14}, 5851 (2022)}.

\bibitem{zhu_topological_2021}
%title = {Topological magnon insulators in two-dimensional van der {Waals} ferromagnets {CrSiTe} $_{\textrm{3}}$ and {CrGeTe} $_{\textrm{3}}$ : {Toward} intrinsic gap-tunability},
F. Zhu, L. Zhang, X. Wang, F. J. Dos Santos, J. Song, \emph{et al}.,
%T. Mueller, K. Schmalzl, W. F. Schmidt, A. Ivanov, J. T. Park, J. Xu, J. Ma, S. Lounis, S. Bl\'{u}gel, Y. Mokrousov, Y. Su, and T. Br\"{u}ckel,
Sci. Adv. \href{https://doi.org/10.1126/sciadv.abi7532}{\textbf{7}, eabi7532 (2021)}.


\bibitem{mermin1966absence}
%  title={Absence of ferromagnetism or antiferromagnetism in one-or two-dimensional isotropic Heisenberg models},
N. D. Mermin and H. Wagner,
Phys. Rev. Lett. \href{https://doi.org/10.1103/PhysRevLett.17.1133}{\textbf{17}, 1133 (1966)}.

\bibitem{tian_magneto-elastic_2016}
%title = {Magneto-elastic coupling in a potential ferromagnetic {2D} atomic crystal},
Y. Tian, M. J. Gray, H. Ji, R. J. Cava, and K. S. Burch,
2D Materials \href{https://doi.org/10.1088/2053-1583/3/2/025035}{\textbf{3}, 025035 (2016)}.


\bibitem{structural-link}
R. Bindu, K. Maiti, S. Khalid, and E. V. Sampathkumaran,
%Structural link to precursor effects,
Phys. Rev. B \href{https://doi.org/10.1103/PhysRevB.79.094103}{\textbf{79}, 094103 (2009)};
%
R. Bindu, G. Adhikary, N. Sahadev, N. P. Lalla, and K. Maiti,
%Pseudogap and charge ordering in a large-bandwidth electron-doped manganite,
Phys. Rev. B \href{https://doi.org/10.1103/PhysRevB.84.052407}{\textbf{84}, 052407 (2011)};
%
D. Lahiri, T. Shibata, S. Chattopadhyay, S. Kanungo, T. Saha-Dasgupta \emph{et al}.,
%R. S. Singh, S. M. Sharma, and K. Maiti,
%Evidence of active role played by the nonmagnetic element Sr in magnetostructural coupling in SrRuO3
Phys. Rev. B \href{https://doi.org/10.1103/PhysRevB.82.094440}{\textbf{82}, 094440 (2010)}.


\bibitem{carteaux_2d_1995}
%title = {{2D} {Ising}-{Like} {Ferromagnetic} {Behaviour} for the {Lamellar} {Cr} $_{\textrm{2}}$ {Si} $_{\textrm{2}}$ {Te} %$_{\textrm{6}}$ {Compound}: {A} {Neutron} {Scattering} {Investigation}},
V. Carteaux, F. Moussa, and M. Spiesser,
EPL \href{https://doi.org/10.1209/0295-5075/29/3/011}{\textbf{29}, 251 (1995)}.


\bibitem{williams2015magnetic}
%  title={Magnetic correlations in the quasi-two-dimensional semiconducting ferromagnet CrSiTe 3},
T. J. Williams, A. A. Aczel, M. D. Lumsden, S. E. Nagler, M. B. Stone, \emph{et al}.,
%J.-Q. Yan, and D. Mandrus,
Phys. Rev. B \href{https://doi.org/10.1103/PhysRevB.92.144404}{\textbf{92}, 144404 (2015)}.

\bibitem{sivadas_magnetic_2015}
%title = {Magnetic ground state of semiconducting transition-metal trichalcogenide monolayers},
N. Sivadas, M. W. Daniels, R. H. Swendsen, S. Okamoto, and D. Xiao,
Phys. Rev. B \href{https://doi.org/10.1103/PhysRevB.91.235425}{\textbf{91}, 235425 (2015)}.


\bibitem{zhang_unveiling_2019}
%title = {Unveiling {Electronic} {Correlation} and the {Ferromagnetic} {Superexchange} {Mechanism} in the van der {Waals} {Crystal} %{CrSiTe} 3},
J. Zhang, X. Cai, W. Xia, A. Liang, J. Huang, \emph{et al}.,
%C. Wang, L. Yang, H. Yuan, Y. Chen, S. Zhang, Y. Guo, Z. Liu, and G. Li,
Phys. Rev. Lett. \href{https://doi.org/10.1103/PhysRevLett.123.047203}{\textbf{123}, 047203 (2019)}.


\bibitem{li_abnormal_2023}
%title = {Abnormal thickness-dependent magneto-transport properties of {vdW} magnetic semiconductor {Cr2Si2Te6}},
Y. Li, Z. Chen, J. Wang, T. Li, M. Tian, J. Karel, and K. Suzuki,
npj 2D Materials and Applications \href{https://doi.org/10.1038/s41699-023-00404-1}{\textbf{7}, 39 (2023)}.


\bibitem{li2019intrinsic}
%  title={Intrinsic van der Waals magnetic materials from bulk to the 2D limit: new frontiers of spintronics},
H. Li, S. Ruan, and Y.-J. Zeng,
Advanced Materials \href{https://doi.org/10.1002/adma.201900065}{\textbf{31}, 1900065 (2019)}.

\bibitem{chen_strain-engineering_2015}
%title = {Strain-engineering of magnetic coupling in two-dimensional magnetic semiconductor {CrSiTe3}: {Competition} of direct exchange interaction and superexchange interaction},
X. Chen, J. Qi, and D. Shi,
Phys. Lett. A \href{https://doi.org/10.1016/j.physleta.2014.10.042}{\textbf{379}, 60 (2015)}.


\bibitem{precursor}
K. Maiti, R. S. Singh, V. R. R. Medicherla, S. Rayaprol, and E. V. Sampathkumaran,
%Origin of Charge Density Wave Formation in Insulators from a High Resolution Photoemission Study of BaIrO$_3$
Phys. Rev. Lett. \href{https://doi.org/10.1103/PhysRevLett.95.016404}{\textbf{95}, 016404 (2005)};
%
P. L. Paulose, N. Mohapatra, and E. V. Sampathkumaran,
%Spin-chain magnetism in Eu-doped Ca$_3$⁢Co$_2$⁢O$_6$ and Ca$_3$⁢CoRhO$_6$ investigated by M\"{o}ssbauer spectroscopy,
Phys. Rev. B \href{https://doi.org/10.1103/PhysRevB.77.172403}{\textbf{77}, 172403 (2008)}.


\bibitem{sengupta2005magnetic}
%  title={Magnetic behavior of Eu Cu 2 As 2: A delicate balance between antiferromagnetic and ferromagnetic order},
K. Sengupta, P. L. Paulose, E. V. Sampathkumaran, Th. Doert, and J. P. F. Jemetio,
Phys. Rev. B  \href{https://doi.org/10.1103/PhysRevB.72.184424}{\textbf{72}, 184424, (2005)}.

\bibitem{pregelj2010magnetic}
%  title={Magnetic phase diagram of the multiferroic FeTe 2 O 5 Br},
M. Pregelj, A. Zorko, O. Zaharko, Z. Kutnjak, M. Jagodi{\v{c}}, \emph{et al}.,
%Z. Jagli{\v{c}}i{\'c}, H. Berger, M. de Souza, C. Balz, M. Lang \emph{et al}.,
Phys. Rev. B \href{https://doi.org/10.1103/PhysRevB.82.144438}{\textbf{82}, 144438 (2010)}.

\bibitem{xie_two_2019}
%title = {Two stage magnetization in van der {Waals} layered {CrXTe3} ({X} = {Si}, {Ge}) single crystals},
Q. Xie, Y. Liu, M. Wu, H. Lu, W. Wang, L. He, and X. Wu,
Materials Letters \href{https://doi.org/10.1016/j.matlet.2019.03.017}{\textbf{246}, 60 (2019)}.


\bibitem{mydosh1993spin}
John A. Mydosh, Spin glasses: an experimental introduction, (CRC Press - 1993).

\bibitem{manganite-spHeat}
R. Bindu, G. Adhikary, S. K. Pandey, S. Patil, and K. Maiti,
%Spectral evolution in an insulator exhibiting linear specific heat
New J. Phys. \href{https://doi.org/10.1088/1367-2630/12/3/033026}{\textbf{12}, 033026 (2010)}.


\bibitem{Ayanesh-SnTe}
% title = {Growth andcharacterization of high-quality single-crystalline SnTe retaining cubic symmetry down to the lowest temperature studied}
A. Maiti, A. Singh, K. K. Iyer, and A. Thamizhavel,
Appl. Phys. Lett. \href{https://doi.org/10.1063/5.0086644}{\textbf{120}, 112102 (2022)}.


\bibitem{kumar2021competing}
%  title={Competing magnetic interactions and magnetoresistance anomalies in cubic intermetallic compounds, Gd 4 RhAl and Tb 4 RhAl, and enhanced magnetocaloric effect for the Tb case},
Ram Kumar, K. K. Iyer, P. L. Paulose, and E. V. Sampathkumaran,
Phys. Rev. Mater. \href{https://doi.org/10.1103/PhysRevMaterials.5.054407}{\textbf{5}, 054407 (2021)}.


\bibitem{milosavljevic_evidence_2018}
%title = {Evidence of spin-phonon coupling in {CrSiTe} 3},
A. Milosavljevi\'{c}, A. \v{S}olaji\'{c}, J. Pe\v{s}i\'{c}, Yu Liu, C. Petrovic, N. Lazarevi\'{c}, and Z. V. Popovi\'{c},
Phys. Rev. B \href{https://doi.org/10.1103/PhysRevB.98.104306}{\textbf{98}, 104306 (2018)}.


\bibitem{sun2018effects}
% title={Effects of hydrostatic pressure on spin-lattice coupling in two-dimensional ferromagnetic Cr2Ge2Te6},
Y. Sun, R. C. Xiao, G. T. Lin, R. R. Zhang, L. S. Ling, \emph{et al}.
%Z. W. Ma, X. Luo, W. J. Lu, Y. P. Sun, and Z. G. Sheng,
Appl. Phys. Lett. \href{https://doi.org/10.1063/1.5016568}{\textbf{112}, 072409 (2018)}.


\bibitem{casto_strong_2015}
%title = {Strong spin-lattice coupling in {CrSiTe} $_{\textrm{3}}$},
L. D. Casto, A. J. Clune, M. O. Yokosuk, J. L. Musfeldt, T. J. Williams, \emph{et al}.
%H. L. Zhuang, M.-W. Lin, K. Xiao, R. G. Hennig, B. C. Sales, J.-Q. Yan, and D. Mandrus,
APL Materials \href{https://doi.org/10.1063/1.4914134}{\textbf{3}, 041515 (2015)}.


\bibitem{lu_lattice_2016}
%title = {Lattice vibrations and {Raman} scattering in two-dimensional layered materials beyond graphene},
X. Lu, X. Luo, J. Zhang, Su Y. Quek, and Q. Xiong,
Nano Research \href{https://doi.org/10.1007/s12274-016-1224-5}{\textbf{9}, 3559 (2016)}.


\bibitem{sun_review_2021}
%title = {Review of {Raman} spectroscopy of two-dimensional magnetic van der {Waals} materials*},
Yu-J. Sun, Si-M. Pang, and J. Zhang,
Chin. Phys. B \href{https://doi.org/10.1088/1674-1056/ac1e0f}{\textbf{30}, 117104 (2021)}.


\end{thebibliography}
\end{document}